%% file: main.tex
\documentclass[letterpaper, 10 pt, conference]{format/ieeeconf}
\IEEEoverridecommandlockouts
\let\labelindent\relax

\usepackage{graphics} 
\usepackage{epsfig} 
\usepackage{times} 
\usepackage{amsmath} 
\usepackage{amsthm}
\usepackage{amssymb}
\usepackage{tikz}
\usetikzlibrary{graphs}
\usepackage[colorlinks=false, urlcolor=blue, pdfborder={0 0 0}]{hyperref}
\usepackage{enumerate}
\usepackage{color}
\usepackage[linesnumbered,boxed,ruled,commentsnumbered]{algorithm2e}
\usepackage{comment}
\usepackage{cite}
\usepackage{tabularx}
\usepackage{booktabs}
\usepackage{amssymb}
\usepackage{subcaption}
\usepackage{enumitem}
\usepackage{booktabs,arydshln}
\newtheorem{mydef}{Definition}
\newtheorem{mythm}{Theorem}
\newtheorem{myprob}{Problem}
\newtheorem{mylem}{Lemma}

\newtheorem{mypro}{Proposition}
\newtheorem{myexm}{Example}
\newtheorem{remark}{Remark}
\newtheorem{assumption}{Assumption}
\usepackage{tablefootnote}
\usepackage[affil-it]{authblk}
\usepackage[framemethod=TikZ]{mdframed}
\usepackage{color,soul}
\mdfsetup{skipabove=5pt,
innertopmargin=0pt}
\DeclareCaptionTextFormat{up}{\MakeTextUppercase{#1}}

\newcommand{\reals}{\mathbb{R}}
\newcommand{\preals}{\reals_{\geq 0}}
\newcommand{\nat}{\mathbb{N}}

\newcommand{\haut}{H}
\newcommand{\hautcontr}{H_I}
\newcommand{\odesol}{\varphi}
\newcommand{\initcond}{x}
\newcommand{\newsafe}{{\mathcal{C}}^*}
\newcommand{\newsafecontrol}{{\mathcal{K}}^*}
\newcommand{\safesw}{\mathcal{S}}
\newcommand{\unsafesw}{\mathcal{U}}
\newcommand{\Cc}{\mathcal{C}}

\title{\LARGE \bf
Safe Control Synthesis for Hybrid Systems through \\ Local Control Barrier Functions
}

\author{
{Shuo Yang\textsuperscript{\rm 1}, Mitchell Black\textsuperscript{\rm 2}, Georgios Fainekos\textsuperscript{\rm 2}, Bardh Hoxha\textsuperscript{\rm 2},
Hideki Okamoto\textsuperscript{\rm 2}, Rahul Mangharam\textsuperscript{\rm 1}}
\thanks{The work was primarily performed while Shuo Yang was with Toyota.}
\thanks{\textsuperscript{\rm 1}University of Pennsylvania; \textsuperscript{\rm 2} Toyota Motor North America, R\&D. Correspondance to:
{\tt\small yangs1@seas.upenn.edu}
}
}

\begin{document}

\maketitle

\begin{abstract}

Control Barrier Functions (CBF) have provided a very versatile framework for the synthesis of safe control architectures for a wide class of nonlinear dynamical systems. 
Typically, CBF-based synthesis approaches apply to systems that exhibit nonlinear -- but smooth -- relationship in the state of the system and linear relationship in the control input.
In contrast, the problem of safe control synthesis using CBF for hybrid dynamical systems, i.e., systems which have a discontinuous relationship in the system state, remains largely unexplored.
In this work, we build upon the progress on CBF-based control to formulate a theory for safe control synthesis for hybrid dynamical systems.
Under the assumption that local CBFs can be synthesized for each mode of operation of the hybrid system, we show how to construct CBF that can guarantee safe switching between modes.
The end result is a switching CBF-based controller which provides global safety guarantees.
The effectiveness of our proposed approach is demonstrated on two simulation studies.
\end{abstract}

\IEEEpeerreviewmaketitle

\input{intro}
\input{problem}

\input{methodology}
\input{experiments}
\input{conclusion}

\bibliographystyle{unsrt}
\bibliography{references}


\end{document}

%% file: intro.tex
\section{Introduction}



Safety-critical control is one of the fundamental problems in autonomous systems. 
Among various safety control methods, control synthesis methods utilizing control barrier functions (CBF) is a recent active area of research.
CBF can explicitly encode  safe sets and enforce the invariance of safe sets via solving efficient online quadratic programs (QP)~\cite{ames2016control, robey2020learning, yang2022differentiable, berducci2023learning}.
A special class of autonomous systems is the class of hybrid dynamical systems, which involves both continuous dynamic flow and discrete dynamical mode jumps for state evolution.
Such discrete mode transitions could be needed to model physical phenomena, or high-level logical decision making.
For instance, vehicle dynamics switching from dry road to wet road could be modeled by a hybrid system.
Bipedal robot walking is another example of a hybrid system.
Safety concerns naturally arise for safety-critical hybrid systems and many safety approaches have been proposed in the past, such as Hamilton-Jacobi reachability-based approaches~\cite{tomlin2000game, bansal2017hamilton} and computing controlled invariant sets~\cite{legat2018computing}.

CBF-based approaches have been used to ensure safety for hybrid systems recently for its computational efficiency~\cite{lindemann2021learning, robey2021learning}, in which the authors use the notion of the global CBF (i.e., all dynamical modes share a common CBF) to ensure safety.
However, such a global notion is restrictive and less necessary since it requires the CBF to satisfy the same invariance conditions for all dynamical modes, which results in the challenge of constructing a valid global CBF, and also in more conservative control behavior.

To mitigate the drawbacks of a global CBF approach, we propose to use multiple local CBFs to guarantee global safety.
We assume that each dynamical mode has its own local CBF, which implies that each mode can be safe under CBF-based control without considering discrete mode switching.
However, it is possible that some \emph{unsafe behaviors can occur under discrete transitions (jumps) even if all modes can be safe independently}.
To ensure global safety of hybrid systems, we first identify those safe and unsafe switching regions.
Then, we refine the initial local CBFs by considering safety after mode switching.
Finally, safety of hybrid systems is guaranteed under refined local CBFs.

\subsection{Related Work}
\paragraph{Safety for hybrid system}
Safety is a paramount concern in hybrid systems and various methods for safety verification and control synthesis have been proposed~\cite{athanasopoulos2018combinatorial, legat2018computing, prajna2004safety, ivanov2019verisig, abate2008probabilistic}. 
Among them, barrier function-based methods can provide provable safety guarantees~\cite{prajna2004safety, glotfelter2019hybrid, maghenem2019characterizations, bisoffi2018hybrid}. 
Furthermore, CBFs emerged as a principled control method to enforce safety for controlled (hybrid) systems~\cite{lindemann2021learning, lavaei2022safety, yang2023safe, maghenem2021adaptive, robey2021learning}.
The most relevant work to ours is~\cite{lindemann2021learning}, in which the authors define a global CBF for hybrid systems and propose a data-driven constructive method to find a valid global CBF.
Our work differs from~\cite{lindemann2021learning} since we focus on ensuring global safety using local CBFs, and we propose to refine local CBFs considering unsafe dynamical mode switching.
Also, we provably show that global CBF is more conservative than local CBFs to gurantee safety for hybrid systems.

\paragraph{Stability for hybrid system}
Our work is also relevant to the Lyapunov-based stability for hybrid systems~\cite{decarlo2000perspectives, chen2021learning}.
As demonstrated in~\cite{liberzon2003switching}, unconstrained switching might lead to global instability even if all dynamical modes are stable with corresponding local Lyapunov functions. 
Thus, finding switching conditions for which global Lyapunov stability is guaranteed is one of the most important and elusive problems in the hybrid systems literature, and many approaches have been proposed~\cite{goebel2009hybrid, branicky1998multiple, hespanha2004uniform, lin2009stability, zhai2006analysis, sun2001control}.
As a dual notion, safety is also of great importance in hybrid systems but its related research is rather limited.
In this paper, we first reveal a similar result that some switching conditions might lead to lack of global safety even if all dynamical modes are safe through control.
Also, we propose an algorithmic procedure to ensure global safety using multiple local CBFs.
To the best of our knowledge, this is the first work addressing the safe switching problem with multiple local CBFs.

The contributions of this paper are summarized below:
\begin{enumerate}
    \item we formulate the safety control problems for hybrid systems using multiple local CBFs;
    \item we reveal that safety might be violated under some switching states even if all dynamical modes are safe through control;
    \item we refine local CBFs by considering safe switching and provide safety guarantees which, to our knowledge, is the first work to do so using multiple local CBFs; and
    \item we demonstrate the effectiveness of our approach through simulations.
\end{enumerate}

%% file: problem.tex
\section{Preliminary and Problem Formulation}
In this section, we cover the background concepts relevant to this work. This includes the related definitions on Control Barrier Functions and Hybrid Systems.
We denote by $\mathbb{R}$ and $\mathbb{R}^n$ the set of real numbers and real $n$-dimensional vectors, respectively. 
The set $\nat$ denotes the natural numbers (including zero). 
We will be using subscripts to denote subsets of these sets, e.g., $\reals_{> 0}$ denotes the set of positive real numbers.
Given a set $X$, $\mathcal{P}(X)$ denotes its powerset.
Let $\alpha$: $\mathbb{R}\rightarrow \mathbb{R}$ denote an extended class $\mathcal{K}_{\infty}$ function, i.e., a strictly increasing function with $\alpha(0)=0$.

\subsection{Control Barrier Functions}
\label{sec:cbf}
Consider a continuous-time
control-affine system: 
\begin{align}\label{eq:system}
    \Dot{x} = f(x) + g(x)u, \quad x(0)=x_0,
\end{align}
where $f$ and $g$ are locally Lipschitz, $x\in D \subseteq \mathbb{R}^n$ is the state and $D$ denotes a compact set in $\mathbb{R}^n$.
Safety can be framed in the context of enforcing set invariance in the state space, i.e., the state should not exit a safe set $\mathcal{C}$. 
The safe set $\mathcal{C}$ is represented by the super-level set
of a continuously differentiable function $h: D \rightarrow \mathbb{R}$. The algebraic expressions for the safe set $\mathcal{C}$ and its boundary $\partial \mathcal{C}$ are given by:
\begin{subequations}
\begin{align}
 \mathcal{C}&=\{x\in D\subset \mathbb{R}^n:h(x)\ge 0\}, \\
 \partial\mathcal{C}&=\{x\in D\subset \mathbb{R}^n:h(x)= 0\}.
\end{align}
\end{subequations}
For a locally Lipschitz continuous control law $u=k(x)$, we have that $\Dot{x}=f(x)+g(x)k(x)$ is locally Lipschitz continuous. Thus, for any initial condition $x_0\in D$, there exists a maximum time interval of existence $I(x_0)=[0,\tau_{max})$, such that $x(t)$ is the unique solution to the ordinary differential equation~(\ref{eq:system}) on $I(x_0)$. We frame the safety of system~\eqref{eq:system} in terms of set invariance as shown below. 
\begin{mydef}(Forward invariance and safety)
The set $\mathcal{C}$ is \emph{forward invariant} if for every $x_0\in \mathcal{C}$, $x(t)\in \mathcal{C}$ holds for all $t\in I(x_0)$. If $\mathcal{C}$ is forward invariant, we say that~\eqref{eq:system} is safe. 
\end{mydef}
To verify invariance of $\mathcal{C}$, a control barrier function can be used as a certificate which characterizes the admissible set of control inputs that render $\mathcal{C}$ forward invariant. 

\begin{mydef}(Control barrier function~\cite{ames2016control})
Let $\mathcal{C}\subset D\subset \mathbb{R}^n$ be the superlevel set of a continuously differentiable function $h: D\rightarrow \mathbb{R}$, then $h$ is
a control barrier function for safe set $\mathcal{C}$ if there exists an extended class $\mathcal{K}_{\infty}$ function $\alpha(\cdot)$ such that for the control system (\ref{eq:system}):
\begin{align}
    \sup_{u \in U}\left[\frac{\partial h(x)}{\partial x}\big(f(x)+g(x)u\big)\right]\ge -\alpha(h(x)),
\end{align}
for all $x\in D$. 
\end{mydef}
Given the CBF $h(x)$, the set of all control values that render $\mathcal{C}$ safe is given by:
\begin{equation}\label{eq:safe_control_set}
    K_{cbf}(x) = \{u\in U: \frac{\partial h(x)}{\partial x}\big[f(x)+g(x)u\big]+\alpha(h(x))\ge 0\}
\end{equation}
which we denote as the safe control set. The following theorem shows that the existence of a control barrier function implies that the control system~\eqref{eq:system} is safe:
\begin{mythm}(\hspace{-0.3pt}\cite[Theorem 2]{ames2016control})
\label{thm:cbf}
Assume $h(x)$ is a CBF on $D \supset \mathcal{C}$ and $\frac{\partial h}{\partial x}(x) \neq 0$ for all $x \in \partial \mathcal{C}$. Then any Lipschitz continuous controller $u(x)$ such that $u(x) \in K_{cbf}(x)$ for all $x \in \mathcal{C}$ will render the set $\mathcal{C}$ forward invariant.
\end{mythm}

\subsection{Hybrid Automaton}
A hybrid automaton is a model of a system with both a continuous dynamic flow and discrete dynamic jumps. 
The state of a hybrid automaton is a pair $(q,x)$ where $q$ is the discrete mode and $x$ is the continuous state vector. 

\begin{mydef}\label{def:hybrid_system}
A hybrid input automaton $\hautcontr$ is a tuple 
$\hautcontr =\langle X, Q, U, U_q, F, 
\texttt{Guard}\rangle$:
\begin{itemize}
    \item $X\subseteq \mathbb{R}^n$ is the continuous state space.
    \item $Q$ is a finite set of modes.
    \item $U\subseteq \mathbb{R}^m$ denotes the continuous space of inputs, and $U_q\subseteq U$ is the admissible control input set for each mode $q\in Q$.
    \item $F: Q\times X\times U_q \rightarrow X$ is a vector field that describes the real-time dynamic flow of the continuous state $x$. 
    For a mode $q\in Q$, we define $F$ as a control affine system with admissible control set $U_q$:
    \begin{equation}
    \label{eq:aff_dyn}
        \Dot{x}=F_q(x, u)=f_q(x)+g_q(x)u.
    \end{equation}
    \item $\texttt{Guard} : Q \times Q \rightarrow \mathcal{P}(X)$ denotes the guard set that triggers mode switching. 
\end{itemize}
\end{mydef}

We only consider deterministic systems in this work, i.e., there is no uncertainty in both the dynamic flow and discrete jumps.
Notice that that Def. \ref{def:hybrid_system} does not allow  jumps in the value of the continuous state of the system.
That is, we assume that the continuous state component of a hybrid system solution is continuous with respect to time.

To provide execution semantics for the hybrid automaton, we will need to define switching feedback control law.

\begin{mydef}
    A switching feedback control law is defined as $k: Q \times X\rightarrow U$, where $k_q(x)$ is locally Lipschitz continuous w.r.t. $x$ for any mode $q$.
\end{mydef}

The composition ($\|$) of a switching feedback controller $k_q$ with a hybrid input automaton $\hautcontr$  will be referred to in the following as a hybrid system. 
We will denote a {\it hybrid system} by $\haut = \hautcontr \| k$.

\begin{mydef}\label{def:hybrid_solutions}
(Hybrid system solution)
For a hybrid system $\haut$ 
and a set of  initial conditions $Q_0 \times X_0\subseteq Q \times X$, a solution (trajectory) of $\haut$ is a sequence $(q_i,\odesol_i,\delta_i)_{i \in N}$, where 
$N$ is $\nat$ or a bounded subset of $\nat$, 
$q_i \in Q$ represents the discrete mode, 
$\odesol_i : X \times \preals \rightarrow X$ represents the continuous state evolution, and 
$\delta_i  \in \preals \cup \{ \infty \}$ represents the duration of operating in mode $i$ (i.e., dwell time), such that 
 \begin{enumerate}
\item $(q_0, \initcond_0)\in Q_0 \times X_0$ is the initial state of the hybrid system at time $\tau_0 = 0$.
\item If there is some $j\in N$ such that 
$\tau_j = \infty$, 
then $j=\max N < \infty$, i.e., $N$ is finite and $j$ is unique.
\item For $i\in N$, we let 
\[ \tau_{i+1} = \sum_{j=0}^{i} \delta_i.\]
If $\tau_i < \infty$ for $i>0$, then $\tau_i$ is the switching time from mode $q_{i-1}$ to $q_{i}$.
\item 
\label{def:hsol:odesol}
For all $i\in N$  and for $t \in [\tau_{i}, \tau_{i+1}]$, $\odesol_i(\initcond_i,t)$ is the solution of (\ref{eq:aff_dyn}) for mode $q_i$ and initial condition $\initcond_i = \odesol_{i-1}(\initcond_{i-1},\tau_{i})$, unless $i=0$ since $\initcond_0$ is defined. 
When $\tau_{i+1}=\infty$, then $t$ ranges over $[\tau_i,\tau_{i+1})$.
\item For all $i\in N$ with $i>0$, if $\tau_{i}<\infty$, then 
$$\odesol_{i-1}(\tau_{i})\in \texttt{Guard}(q_{i-1}, q_{i}).$$
\end{enumerate}
%
%
\end{mydef}

The above conditions require that a mode transition happens when the continuous state belongs to the $\texttt{Guard}$ set.
Then, the system follows the continuous flow of the new mode until the next mode transition occurs.

In the following, we will denote by $\mathcal{L}_{\haut}(Q_0,X_0)$ the set containing all solutions of $\haut$ with initial conditions $Q_0 \times X_0$.
If $(Q_0,X_0) = (Q,X)$, i.e., any initial condition is possible, then we just write $\mathcal{L}_{\haut}$ for the language.
We will also use the notation $q \rightarrow q'$ to represent the transition $(q, q')$ when $\texttt{Guard}(q,q')\neq \emptyset$.

Notice that in Def. \ref{def:hybrid_solutions}, we do not explicitly impose any conditions on the input signal $u$.
However, Def. \ref{def:hybrid_solutions}.\ref{def:hsol:odesol} requires that a solution to ($\ref{eq:aff_dyn}$) exists.
In addition, condition \ref{def:hsol:odesol} enforces continuity of the continuous state vector at discrete mode transition times, i.e., $\odesol_{i+1}(\initcond_{i+1}, \tau_{i+1}) = \initcond_{i+1} = \odesol_i(\initcond_i,\tau_{i+1})$.
Therefore, in the following, we can view the solution of the hybrid system as a function of time, i.e., $x : \preals \rightarrow X$, and ignore the jump index $i$, or hybrid mode $q$ when they are not important.

\subsection{Problem Formulation}


This paper is concerned about the safety of the hybrid system in Def.~\ref{def:hybrid_system}.
Before we formulate the problem statement, we first introduce an illustrative example.

\begin{figure*}[!t]
     \centering
     \begin{subfigure}{1.4 \columnwidth}
        \centering
        \includegraphics[width=\textwidth]{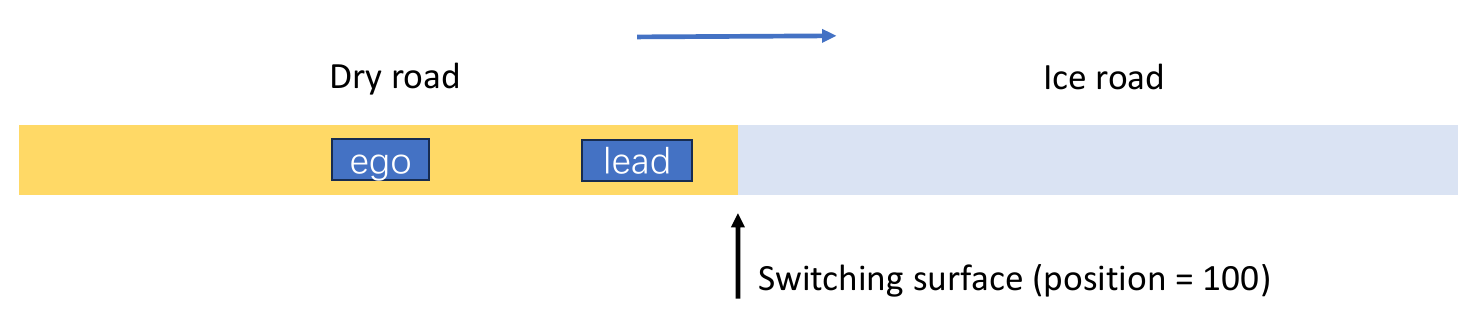}
        \caption{Adaptive cruise control scenario. There is a jump between dry road and ice road, and the ego car is expected to follow the leading car while maintaining a safe distance. Intuitively, the ego car cannot switch with a very high speed.}
        \label{fig:acc-scenario}
     \end{subfigure}
     \hfill
     \begin{subfigure}{1.4 \columnwidth}
        \centering
        \includegraphics[width=\textwidth]{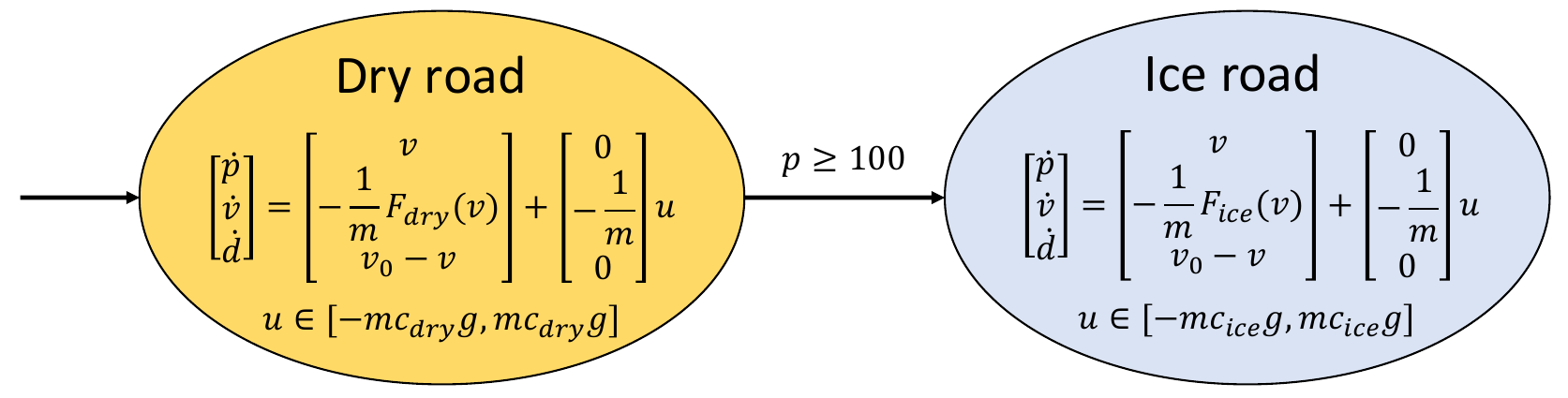}
        \caption{Hybrid dynamics for adaptive cruise control. Please refer to Sec.~\ref{sec:experiment-acc} for dynamics details.}
        \label{fig:acc-dynamics}
     \end{subfigure}
            \caption{Hybrid adaptive cruise control.}
        \label{fig:acc}
\end{figure*}

\begin{myexm}\label{exm:acc}
    Let us consider an adaptive cruise control system.
    As shown in Figure~(\ref{fig:acc-scenario}), the ego car and the leading car are driving on a straight road. The ego car is required to follow the leading and maintain a safe distance.
    The road is partitioned into dry road and ice road with different frictions and control input bounds.
    So there are two modes of dynamics in this example.
    The hybrid system model is shown in Fig.~(\ref{fig:acc-dynamics}).
    The guard set $\texttt{Guard}$ for the transition from ``dry road" to ``ice road" contains all states whose position $p$ of the ego car is greater than $100m$.
    
    Intuitively, the ego car should avoid high speed while switching from dry road to ice road, since both the friction and control bound of the ice road dynamics are smaller so the ego might not be able to decrease the speed as fast as in dry road.
    Unsafe behavior can occur after switching from dry to ice even if the ego was safe on the dry road.
    This implies that having two CBFs, one for dry and one for ice road dynamics, and applying them as safety filters is not sufficient for global safety guarantees.
    Therefore, we are aiming to study the safety when considering  switching dynamics.
\end{myexm}

We first define $(q, q')$-safety for hybrid systems, and then define global safety.

\begin{mydef}
    ($(q, q')$-safety for hybrid system) 
    For a hybrid system $\haut$, a pair of modes $(q,q') \in Q^2$, and safe sets $\mathcal{C}_q, \mathcal{C}_{q'}\subset D$ for modes $q$ and $q'$, respectively,  
    we say that $\haut$ is $(q, q')$-safe w.r.t. $\mathcal{C}_q$ and $\mathcal{C}_{q'}$ 
    if for any initial state $(q_0,\initcond_0)$ with $q_0=q$ and $\initcond_0\in\Cc_q$, the (potentially bounded) resulting trajectory 
    $(q_i,\odesol_i,\delta_i)_{i\in\{0,1\}}$ of $\haut$ with $q_1 = q'$ satisfies:
    \begin{itemize}
        \item $\odesol_0(\initcond_0,t) \in \mathcal{C}_q$ for all $t \in [\tau_0, \tau_1]$, and
        \item $\odesol_1(\initcond_1,t) \in \mathcal{C}_{q'}$ for all $t \in [\tau_1,\tau_2]$ if $\delta_1<\infty$ or for all $t \geq \tau_1$ otherwise.
    \end{itemize}
\end{mydef}

The above $(q, q')$-safety definition  enforces safety requirements for any trajectory transitioning from mode $q$ to $q'$. 
Note that the safe set $\mathcal{C}_q$ for flow mode $q$ and $\mathcal{C}_{q'}$ for $q'$ will be different in general.
Nevertheless, for $(q, q')$-safety to hold (to be enforceable) it must be the case that $\mathcal{C}_q \cap \mathcal{C}_{q'} \cap \texttt{Guard}(q,q') \neq \emptyset$.

We define the set of possible mode transition pairs of a hybrid system $\haut$ as
\[
    \mathcal{T}(\haut) = \{(q_i, q_{i+1})_{i \in N\backslash \sup N}\; |\;
    (q_i,\odesol_i,\delta_i)_{i \in N}\!\in\!\mathcal{L}_{\haut}\}.
\]
We can now define global safety based on $(q, q')$-safety.

\begin{mydef}
\label{def:global:multi:cbf}
    (Global safety for hybrid system) For a hybrid system $\haut$, for given safe sets $\mathcal{C}_q\subset D$ for any mode $q\in Q$, we say that $\haut$ is globally safe w.r.t. $\{\mathcal{C}_q\}_{q\in Q}$ if $\haut$ is $(q, q')$-safe for any $(q, q')\in \mathcal{T}(\haut)$.
\end{mydef}

Similar to $(q, q')$-safety, $\mathcal{C}_q \cap \mathcal{C}_{q'} \cap \texttt{Guard}(q,q') \neq \emptyset$ must also hold for any $(q, q')\in \mathcal{T}(H)$ to ensure that global safety is enforceable.
In the following, we will assume that the safe sets for each mode of the hybrid automaton are provided since our goal is to synthesize a switching controller that guarantees global safety.

\begin{assumption}\label{assumption:multiple-cbfs}
    For any mode $q\in Q$ of $\hautcontr$, we assume that we have a local control barrier function $h_q$ for 
    (\ref{eq:aff_dyn})
    and the corresponding safe set $\mathcal{C}_q=\{x\in D\subset\mathbb{R}^n: h_q(x)\ge 0\}$.
For every $q\in Q$, we denote its safe control set by 
    \[
    K_q(x)\!=\!\{u\!\in\! U_q:\!\frac{\partial h_q(x)}{\partial x}\big[f_q(x)+g_q(x)u\big]\ge -\alpha_q(h_q(x))\},
    \]
    where $\alpha_q$ is the corresponding extended class $\mathcal{K}_{\infty}$ function.
\end{assumption}

The above assumption says that every mode of the hybrid automaton is equipped with its own local CBF.
This is a mild assumption given the growing literature on the synthesis of CBF~\cite{xiao2023safe, clark2021verification, wang2018permissive, robey2020learning}.
However, local CBF cannot guarantee safety when switching between different modes.
In other words, applying a control input $u\in K_q$ when the mode is $q$ is not enough to ensure safety when the system switches to a mode $q'$.
Thus, we formulate the following problem which is addressed in this paper.

\begin{myprob}
    Given a hybrid automaton $\hautcontr$, under Assumption~\ref{assumption:multiple-cbfs}, we are interested in solving two sub-problems:
    \begin{enumerate}
        \item find a switching control law $k$ that can ensure the $(q, q')$-safety of $\haut$, where $q, q'\in Q$;
        \item find a switching control law $k$ that can ensure the global safety of $\haut$.
    \end{enumerate}
    
\end{myprob}

To solve the above problem, we first identify safe and unsafe switching sets, and then compute the unsafe backward reachable set. 
Finally, we refine the initial local CBFs by avoiding the new unsafe sets. 
The refined CBFs can guarantee safety for the hybrid system. 

%% file: methodology.tex
\section{Safe control for hybrid systems}

In this section, we formalize our notions of safety for hybrid dynamical systems, and we provide sufficient conditions for safe control synthesis.
We start by defining what safe and unsafe switching sets are.

\begin{mydef}\label{def:switching-set}
    (Safe and unsafe switching sets) For any mode jump $q\rightarrow q'$ in $\haut$, the corresponding safe switching set is defined by $\safesw_{q, q'}=\texttt{Guard}(q, q')\cap\mathcal{C}_q\cap\mathcal{C}_{q'}$ and the corresponding unsafe switching set is defined by $\unsafesw_{q, q'}=(\texttt{Guard}(q, q')\cap\mathcal{C}_q)\symbol{92}\safesw_{q, q'}$.
\end{mydef}

Safety can be preserved when the switching state is in the safe switching set $\safesw_{q, q'}$.
However, safety is jeopardized when the switching state is in the unsafe switching set.
This intuition is formalized below. 

\begin{mypro}\label{prop:switch}
     For a hybrid system $\haut$
     and a given initial condition $(q_0,\initcond_0)$, for any (potentially bounded) trajectory 
     $(q_i,\odesol_i,\delta_i)_{i \in \{0, 1\}}$ 
     of $\haut$ that satisfies $q_0=q$ and $q_1=q'$, 
     we have that:
     \begin{itemize}
         \item if $\odesol_0(\initcond_0,t)\in \mathcal{C}_q$ for all $t\in [\tau_0, \tau_1]$, and $\odesol_0(\initcond_0,\tau_1)\in \safesw_{q, q'}$, then $\odesol_1(\initcond_1,t)\in \mathcal{C}_{q'}$ under $k_{q'}(\odesol_1(\initcond_1,t))\in K_{q'}(\odesol_1(\initcond_1,t))$  for all $t \in[\tau_1, \tau_2]$, i.e., $\haut$ is $(q, q')$-safe;
         \item if $\odesol_0(\initcond_0,t)\in \mathcal{C}_q$ for all $t\in [\tau_0, \tau_1]$ and $\odesol_0(\initcond_0,\tau_1)\in \unsafesw_{q, q'}$, then $\haut$ is not $(q, q')$-safe.
     \end{itemize}
        
\end{mypro}
\begin{proof}
    First, if the switching state $\odesol_0(\initcond_0,\delta_0)\in \safesw_{q, q'}=\texttt{Guard}(q, q')\cap\mathcal{C}_q\cap\mathcal{C}_{q'}$, then we have $\odesol_0(\initcond_0,\delta_0)\in \mathcal{C}_{q'}$.
    Since we already know that $\mathcal{C}_{q'}$ is forward invariant under the safe control set $K_{q'}$, then we know that $\odesol_1(\odesol_0(\initcond_0,\delta_0),t)\in \mathcal{C}_{q'}$ can hold for all $t\in [\tau_1, \tau_2]$.
    Second, however, if $\odesol_0(\initcond_0,\tau_1)\in \unsafesw_{q, q'}$, this means that the switching state is not in the safe set $\mathcal{C}_{q'}$, which directly implies that $\haut$ is not $(q, q')$-safe.
\end{proof}

The above proposition states that the continuous state must be in the safe switching set to ensure the $(q, q')$-safety.
It also implies that the safety of a hybrid system  can still be violated even if each mode is safe under the corresponding CBF safety filter.
Next, Algorithm~\ref{alg:procedure} is provided to guide the system state to reach the safe switching set and avoid 
 the unsafe switching set when the system is must switching.



\begin{algorithm}[!h]
    \SetAlgoLined
    \LinesNumbered
    \caption{Procedure for $(q, q')$-safety synthesis}
    \label{alg:procedure}
    
    \SetKwInOut{Input}{Input}
    \SetKwInOut{Output}{Output}
    
    
    Identify the safe switching set $\safesw_{q, q'}$ and the unsafe switching set $\unsafesw_{q, q'}$ for each $q\rightarrow q'$\;
    Compute the backward reachable set $\texttt{BackUnsafe}_{q, q'}$ for the unsafe switching set $\unsafesw_{q, q'}$\;
    Obtain the new CBF $h_{q, q'}$ for $q\rightarrow q'$ by refining the initial CBF of mode $q$ (i.e., $h_q$) via considering the backward unsafe set using dynamic programming\;
    Control the system with $h_{q, q'}$ when system has mode $q$, and with $h_{q'}$ when system has mode $q'$\;
    
\end{algorithm}


Step 1 of Alg.~\ref{alg:procedure} is typically easy to compute depending on the computational representation of the sets, i.e., safe sets and guard sets.
A polyhedral set representation is closed under intersection, union and complementation and it is typically used when modeling guards in hybrid systems \cite{AlthoffFG2021review}.
However, the results on the synthesis of CBF-based controllers with polyhedral safety sets are limited \cite{thirugnanam2022safety}.
Therefore, set under-approximations using ellipsoidal sets will typically need to be computed \cite{BoydV_book04}.

Next, we discuss the computation of the backward reachable sets of unsafe guard conditions in Step 2.
Let $C(U)$ be the set of all functions from positive reals to some set $U$, i.e., $C(U) = U^{\preals}$.

\begin{mydef}
    (Unsafe backward set) For any jump $q\rightarrow q'$ in $\hautcontr$, the corresponding unsafe backward set is defined by 
    \begin{align}
        \texttt{BackUnsafe}_{q, q'}= & \{x_0 \in C_q \; | \; \forall u \in C(U_q), \exists T \in \preals, \nonumber\\ 
        & \text{s.t. } \odesol_q(x_0, T) \in \unsafesw_{q, q'} \}.
    \end{align}
\end{mydef}

$\texttt{BackUnsafe}_{q, q'}$ contains all states that will inevitably enter the unsafe switching set $\unsafesw_{q, q'}$ in finite time no matter what control signal is applied.
Therefore, we need to control the system to avoid the unsafe backward set, otherwise it will definitely enter $\unsafesw_{q, q'}$ and invalidate safety.
There are some approaches to compute the backwards reachable set of the given target set, see, e.g., ~\cite{bansal2017hamilton, kurzhanski2000ellipsoidal, dreossi2016parallelotope, majumdar2014convex}. We use Hamilton-Jacobi reachability in this paper to compute $\texttt{BackUnsafe}_{q, q'}$. The readers are refered to~\cite{bansal2017hamilton} for technical details of Hamilton-Jacobi reachability.

\begin{remark}
    In general, the set $\mathcal{C}_q\symbol{92}\texttt{BackUnsafe}_{q, q'}$ is not a controlled invariant set\footnote{A set $C$ is called a controlled invariant set if any trajectory starting within $C$ can always be controlled to remain inside $C$. For example, the superlevel set of a CBF is a controlled invariant set.}. This motivates us to find a new CBF $h_{q, q'}$ such that for the new safe set 
    \[ \mathcal{C}_{q, q'}=\{x\in D\subset \mathbb{R}^n \; | \; h_{q, q'}(x)\ge 0\},\] 
    we have that $\mathcal{C}_{q, q'}\subseteq\mathcal{C}_q\symbol{92}\texttt{BackUnsafe}_{q, q'}$.
    Here, we find the new CBF $h_{q, q'}$ by refining the initial local CBF $h_{q}$.
\end{remark}

Next, we introduce how to find the new control barrier function $h_{q, q'}$ using dynamic programming (Step 3 of Alg.~\ref{alg:procedure}).
By leveraging techniques from~\cite{tonkens2022refining}, we update CBF $h_{q}$ recursively using Hamilton-Jacobi reachability.
When the process terminates, we obtain a valid CBF $h_{q, q'}$ on $\mathcal{C}_q\symbol{92}\texttt{BackUnsafe}_{q, q'}$. 
The new safe set $\mathcal{C}_{q, q'}$ is the superlevel set of $h_{q, q'}$.
The validity of $h_{q, q'}$ is established in the following results.

\begin{mylem}
    (Adapting Theorem 1 from~\cite{tonkens2022refining}) The refined CBF $h_{q, q'}$ is valid upon convergence, i.e., there exists an extended class $\mathcal{K}_{\infty}$ function $\alpha_{q, q'}(\cdot)$ such that:
\begin{align}
    \mathsf{sup}_{u\in U_q}\frac{\partial h_{q, q'}(x)}{\partial x}[f_q(x)\!+\!g_q(x)u]\!\ge\! -\alpha_{q, q'}(h_{q, q'}(x)),
\end{align}
for all $x\in \mathcal{C}_{q, q'}$.
\end{mylem}

Our initial $h_{q}$ can be considered as a good warmstarting for the CBF refinement, which can accelerate the convergence of the recursive update.
However, note that this dynamic programming (DP)-based CBF refinement is generally limited to low-dimensional systems~\cite{tonkens2022refining}. 
This is typically the case since   spatially discretized DP recursion results in exponential computational complexity w.r.t. the system dimensionality.

\begin{remark}
    From~\cite{tonkens2022refining,choi2021robust}, we know that the converged CBF recovers a valid control barrier-value function (CBVF), and that CBVF recovers the largest controlled invariant set.
    Hence, our refined CBF-based  method is not conservative under the given multiple local CBFs.
\end{remark}

Then, we can obtain the $(q, q')$-safety guarantees. 
Using notation similar to~(\ref{eq:safe_control_set}), $K_{q, q'}(x)$ denotes the safe control set defined by  $h_{q, q'}$.


\begin{mythm}\label{thm:switch-safety}
    For any initial state $x_0\in \mathcal{C}_{q, q'}$ at mode $q$, a hybrid system $\haut$ is $(q, q')$-safe under any switching feedback controller $k$ s.t. $k_q(x) \in K_{q, q'}(x)$ and $k_{q'}(x) \in K_{q'}(x)$.
\end{mythm}

\begin{proof}
    Consider any trajectory $(q_i,\odesol_i,\delta_i)_{i \in \{0, 1\}}$ of $\haut$ satisfying $q_0=q$ and $q_1=q'$.
    For any $x \in \mathcal{C}_{q, q'}$, any control input $k_q(x) \in  K_{q, q'}(x)$ maintains the forward invariance of the safe set $\mathcal{C}_{q, q'}$.
    Thus,  $\odesol_0(x_0,t)\in\mathcal{C}_{q, q'}\subseteq \mathcal{C}_q$ for all $t\in [0, \delta_0]$.
    Now, according to Prop.~\ref{prop:switch}, we only need to prove that the switching state is in the safe switching set, i.e.,  $\odesol_0(x_0,\delta_0) \in \safesw_{q, q'}$.
    Since $\odesol_0(x_0,\delta_0) \in\mathcal{C}_{q, q'}$ and $\mathcal{C}_{q, q'}\subseteq\mathcal{C}_q\symbol{92}\texttt{BackUnsafe}_{q, q'}$, then $\odesol_0(x_0,\delta_0) \not\in\texttt{BackUnsafe}_{q, q'}$.
    Also, since $\unsafesw_{q, q'}\subseteq\texttt{BackUnsafe}_{q, q'}$, then we have that $\odesol_0(x_0,\delta_0) \not\in \unsafesw_{q, q'}$.
    Note that $\odesol_0(x_0,\delta_0) \in \texttt{Guard}_{q, q'}\cap \mathcal{C}_{q, q'}\subseteq \texttt{Guard}_{q, q'}\cap \mathcal{C}_{q} = \safesw_{q, q'}\cup \unsafesw_{q, q'}$.
    Hence, we finally obtain $\odesol_0(x_0,\delta_0) \in \safesw_{q, q'}$.
\end{proof}

\begin{remark}
    Note that when $\mathcal{C}_{q, q'}\cap \safesw_{q, q'}=\emptyset$, then there is no safe switching from mode $q$ to mode $q'$.
    In this case, the safe control input set $K_{q, q'}$ will prevent the continuous state from entering the guard set and switching mode.
    Also, when the initial state is in $\mathcal{C}_q\symbol{92}\mathcal{C}_{q, q'}$, then the system cannot be $(q, q')$-safe, since the 0-superlevel set of $h(q, q')$ has already recovered the largest controlled invariant set~\cite{choi2021robust, tonkens2022refining}.
\end{remark}

After deriving the safety control method for $(q, q')$-safety, we can now present our main result on global safety.
Given a mode $q$, we define a new safe set $\newsafe_{q} = \bigcap_{(q, q')\in \mathcal{T}(\haut)}\mathcal{C}_{q, q'}$ which is the intersection of all safe sets for the safe jump to any next possible mode.
The resulting safe control set $\newsafecontrol_{q}(x)=\bigcap_{(q, q')\in \mathcal{T}(\haut)}K_{q, q'}(x)$ is the intersection of all safe control sets for safe transition to any next possible mode.

\begin{mythm}\label{thm:global-safety}
    Under the assumptions that
    \begin{enumerate}
        \item For the given set of initial conditions $Q_0 \times X_0\subseteq Q \times X$, we have  $x_0 \in \newsafe_{q_0}  \not=\emptyset$ for any initial state $(q_0,x_0) \in Q_0 \times X_0$; and
        \item $\forall (q, q')\in \mathcal{T}(\haut): \newsafe_{q}\cap \newsafe_{q'}\cap \texttt{Guard}(q, q')\not=\emptyset$ or $\newsafe_{q}\cap \texttt{Guard}(q, q') = \emptyset$; and
        \item $\forall q\in Q, \forall x\in \newsafe_{q}: \newsafecontrol_{q}(x)\not=\emptyset$ if $\newsafe_{q}\not=\emptyset$,
    \end{enumerate}
    then for any initial state $(q_0,x_0) \in Q_0 \times X_0$, the hybrid system $\haut$ is globally safe under any switching $k$ control which satisfies $k_{q}(x(t))\in \newsafecontrol_{q}(x(t))$.  
\end{mythm}
\begin{proof}
    Consider any trajectory $(q_i,\odesol_i,\delta_i)_{i \in N}$ of $\haut$.
    For any $i \in N\backslash \sup N$ and any feedback switching control $k_{q_i}(\odesol_i(x_i, t))\in \newsafecontrol_{q_i}(\odesol_i(x_i, t))$, since $\newsafecontrol_{q_i}(\odesol_i(x_i, t))\subseteq K_{q_i, q_{i+1}}(\odesol_i(x_i, t))$, so $k_{q_i}(\odesol_i(x_i, t))\in K_{q_i, q_{i+1}}(\odesol_i(x_i, t))$; also, $k_{q_{i+1}}(\odesol_{i+1}(x_{i+1}, t))\in  \newsafecontrol_{q_{i+1}}(\odesol_{i+1}(x_{i+1}, t))$ ensures that $\odesol_{i+1}(x_{i+1}, t)\in \newsafe_{q_{i+1}}\subseteq \mathcal{C}_{q_{i+1}}$. Thus, $(q_i, q_{i+1})$-safety is guaranteed according to Theorem~\ref{thm:switch-safety} and $\haut$ is globally safe.
    
    The first assumption says the initial state should be safe. For the second assumption, it is necessary because the case $\newsafe_{q}\cap \newsafe_{q'}\cap \texttt{Guard}(q, q')\not=\emptyset$ means that safe switching set (see Def.~\ref{def:switching-set}) is not empty, which make the safe switching feasible (see Prop.~\ref{prop:switch}). On the other hand, if the safe switching set is empty, we can also ensure safety as long as $\newsafe_{q}\cap \texttt{Guard}(q, q') = \emptyset$, i.e., the switching cannot happen, which is the latter case of the second assumption.
    
For the third assumption, we do not assume $\newsafecontrol_{q}\not=\emptyset$ if $\newsafe_{q}=\emptyset$. We illustrate its reason by proving that mode $q$ will not appear in any trajectory if $\newsafe_{q}=\emptyset$, i.e.,  
there exists no $i \in N\backslash \sup N$ such that $q_{i+1}=q$.
Suppose there exists such $i \in N\backslash \sup N$ with $q_{i+1}=q$, then $\newsafe_{q_i}\cap \newsafe_{q_{i+1}}\cap \texttt{Guard}(q_i, q_{i+1})=\emptyset$ since $\newsafe_{q_{i+1}}=\emptyset$. 
Also, $\newsafe_{q_i}\cap \texttt{Guard}(q_i, q_{i+1}) \not = \emptyset$ holds, otherwise the transition from mode $q_i$ to mode $q_{i+1}$ is impossible under switching control $k_{q_i}(\odesol_i(x_i, t))\in \newsafecontrol_{q_i}(\odesol_i(x_i, t))$.
Therefore, the second assumption cannot hold, which is a contradiction. 
Thus, mode $q$ will not be in any trajectory if $\newsafecontrol_{q}\not=\emptyset$, so we do not need to pose any requirement on $\newsafecontrol_{q}$.
\end{proof}

\begin{figure}[h]
    \centering
    \includegraphics[width=0.4\textwidth]{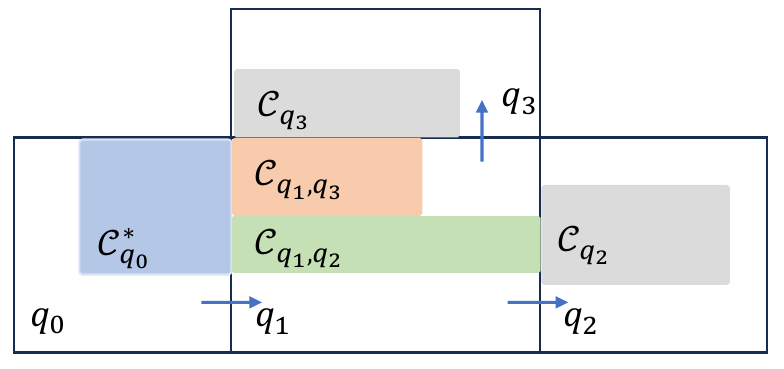}
    \caption{One hybrid automaton example where safe switching controller exists but Theorem~\ref{thm:global-safety} is not applicable.}
    \label{fig:counter-ex}
\end{figure}

Note that the above theorem provides \emph{sufficient but not necessary} conditions for safe feedback switching controller design. 
There might exist a safe control law even if the conditions of Theorem~\ref{thm:global-safety} are not satisfied. 
For example, consider the hybrid automaton in Fig.~\ref{fig:counter-ex}, where the transitions are denoted by blue arrows and the $\texttt{Guard}$ set for each transition is the common edge between  two modes. 
The hybrid system does not satisfy the second assumption of Theorem~\ref{thm:global-safety} since $\newsafe_{q_1}=\emptyset$ and $\newsafe_{q_0}\cap \texttt{Guard}(q_0, q_1) \not= \emptyset$.
However, when the trajectory starting from $\newsafe_{q_0}$ is switching from mode $q_0$ to $q_1$, no matter whether it ends in $\mathcal{C}_{q_1, q_2}$ or $\mathcal{C}_{q_1, q_3}$, the system  can still be safely controlled toward the next possible switch ($q_1\rightarrow q_2$ or $q_1\rightarrow q_3$).

\begin{remark}
    It is possible that the above multiple CBF constraints sometimes lead to infeasibility, i.e., $\newsafecontrol_{q}(x(t))=\emptyset$. 
    This problem can be addressed by considering a subset of feasible transitions, or even by staying in a specific mode if this mode can guarantee safety without any jumps. 
    
    In many cases, e.g., in automated driving systems, we may only need to consider one next dynamics mode transition, e.g., based on path planning or prediction, so one CBF constraint is sufficient to ensure safety for each mode transition (i.e., global safety reduces to sequential $(q, q')$-safety).
    Theoretically, improving feasibility under multiple CBFs is an important problem and has been recenlty addressed using different approaches~\cite{tan2022compatibility, breeden2023compositions, black2023consolidated}.
\end{remark}

One may ask about the relationship between global hybrid system safety (Def. \ref{def:global:multi:cbf}) through multiple local CBFs and safety induced through a global CBF. 
In the following proposition, we show that the local CBFs-based method is generally less conservative than global safety guaranteed by a global CBF-based method. 
We first formally define global CBF for hybrid systems.

\begin{mydef}
    $h_g(x)$ is a global CBF for the hybrid automaton $\hautcontr$  if there exists an extended class $\mathcal{K}_{\infty}$ function $\alpha(\cdot)$ such that for $\hautcontr$:
\begin{align}\label{CBF-condition}
    \mathsf{sup}_{u\in U_q}\frac{\partial h_g(x)}{\partial x}[f_q(x)+g_q(x)u]\ge -\alpha(h_g(x))
\end{align}
holds for all $(q, x)\in Q\times X$. 
\end{mydef}

\begin{mypro}
    If there exists a global CBF $h_g$ for the hybrid automaton $\hautcontr$ for its corresponding safe set $\mathcal{C}_g$, then there exist a local CBF $h_q$ for each mode $q\in Q$ such that the state can always stay inside $\mathcal{C}_g$.
\end{mypro}
\begin{proof}
    We can let $h_q=h_g$ for each mode $q\in Q$.
\end{proof}

The above proposition establishes that a global CBF can be viewed as a special case of the local CBFs method. 
To further highlight the differences, we make several observations: 
\begin{itemize}
    \item Global CBF asks for the same safe state set for each flow mode, 
    but the local CBFs-based method can support different safe sets in each mode.
    Therefore, control synthesis for global safety through local CBF can support a wider range of safe control applications. 
    \item The single global CBF should satisfy constraints for all flow modes, but any local CBF is only responsible for its own specific mode. This implies that global CBF are harder to synthesize than local CBFs.
    \item Global CBF impose more restrictive constraint conditions than local CBFs to enforce safety because of the previous observation. This means that local CBFs can lead to better system performance while ensuring safety.
\end{itemize}



%% file: experiments.tex
\section{Case studies}
In this section, we present two case studies to illustrate the effectiveness of our multiple local CBFs-based safe control method. 
We compare our approach with a baseline approach where each initial local CBF is applied for each dynamical mode and it is unaware of the safety effects of modes switching (we name this approach briefly as switch-unaware CBF). We also compare ours with global CBF method.

\subsection{Adaptive cruise control}\label{sec:experiment-acc}
We first conduct simulation on the aforementioned adaptive cruise control example.
As described in Example~\ref{exm:acc}, the road has two different surfaces (dry road and ice road) with different dynamical modes.
Hence, two different local CBFs are required (and assumed) to ensure safety for each mode.
The system state is $x=[p\; v\; d]^T$, where $p$ is the ego car position, $v$ is the ego car velocity, and $d$ is the distance between the two cars. 
The control input $u$ is the acceleration.
The switching scenario is presented in Fig.~\ref{fig:acc-scenario} and the hybrid model is shown in Fig.~\ref{fig:acc-dynamics}, where 
$m$ is the mass of the vehicle, $g$ is the gravitational constant, $v_0$ is the velocity of leading car, $c_{dry}$ ($c_{ice}$) is the maximum $g$-force that can be applied on dry (ice) road, and $Fr_{dry}(v) = f_{0, d}v^2 + f_{1, d}v + f_{2, d}$ is the friction of dry road (correspondingly, $Fr_{ice}(v)$ is the friction of ice road but with different parameters).

\begin{figure}[!tbp]
  \centering
    \begin{subfigure}[b]{0.157\textwidth}
    \includegraphics[width=\textwidth]{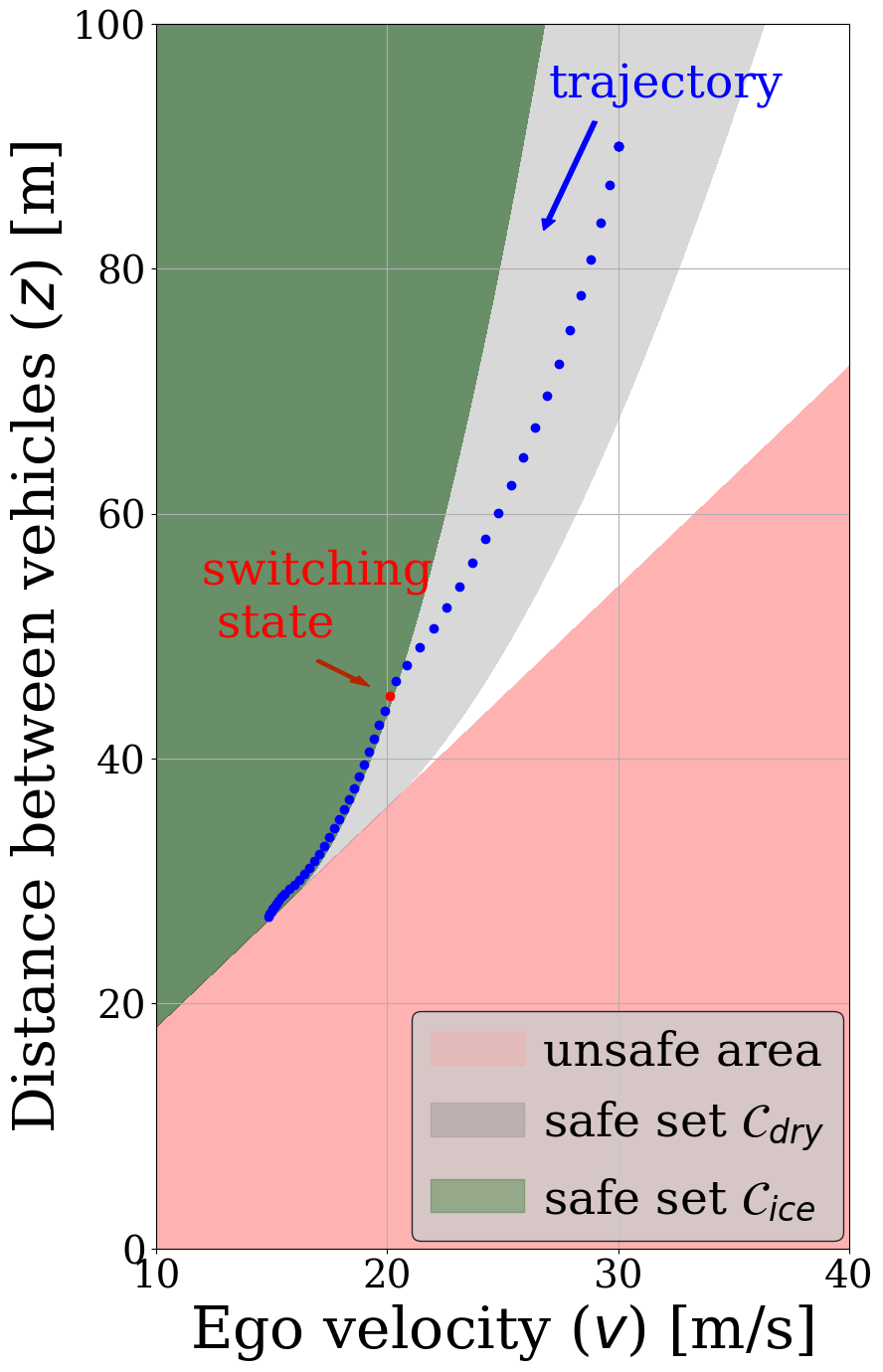}
    \caption{\scriptsize{Switch-aware CBF}}
    \label{fig:ours-good-final}
  \end{subfigure}
  \begin{subfigure}[b]{0.157\textwidth}
    \includegraphics[width=\textwidth]{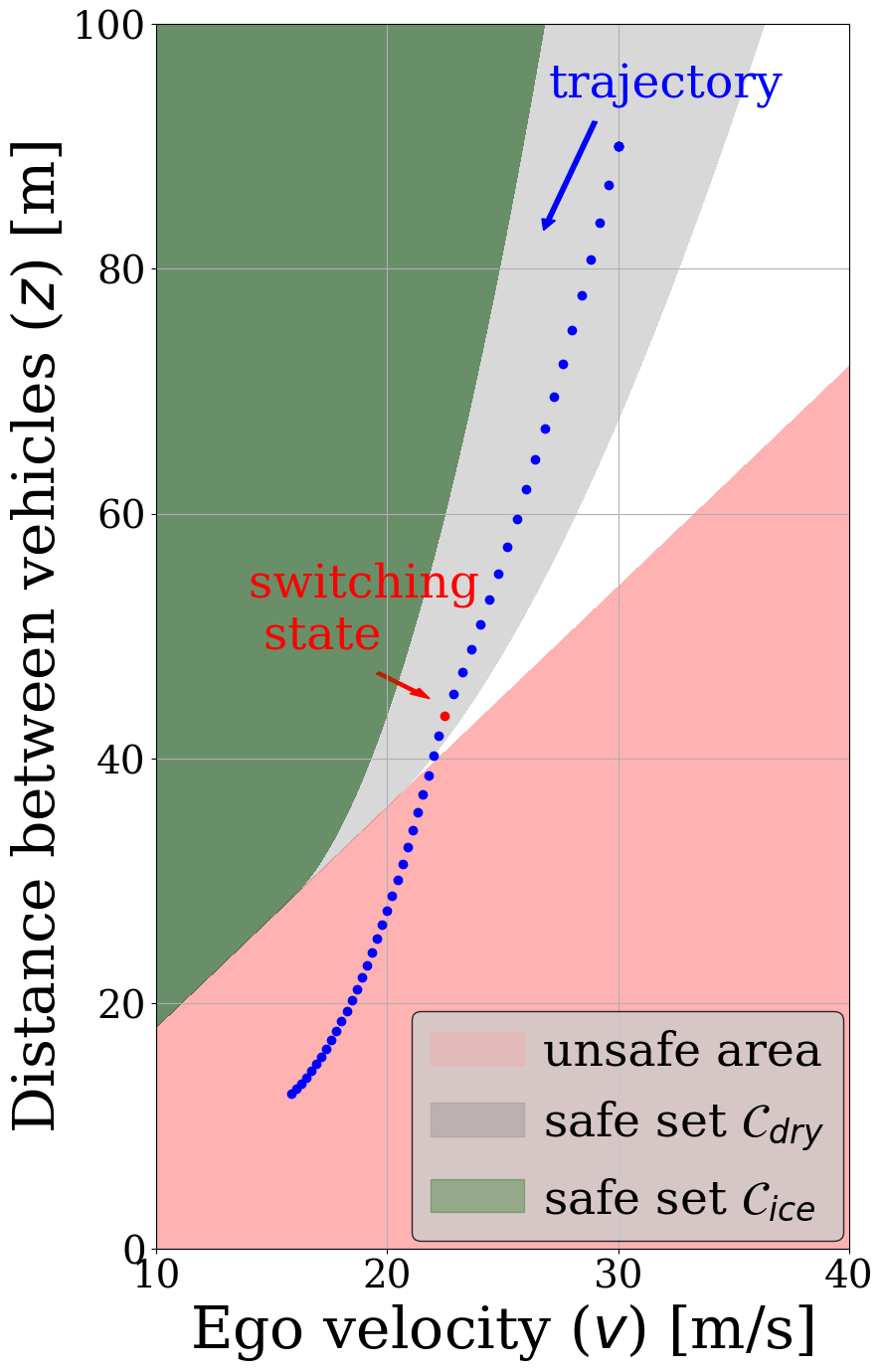}
    \caption{\scriptsize{Switch-unaware CBF}}
    \label{fig:baseline-bad-final}
  \end{subfigure}
  \begin{subfigure}[b]{0.157\textwidth}
    \includegraphics[width=\textwidth]{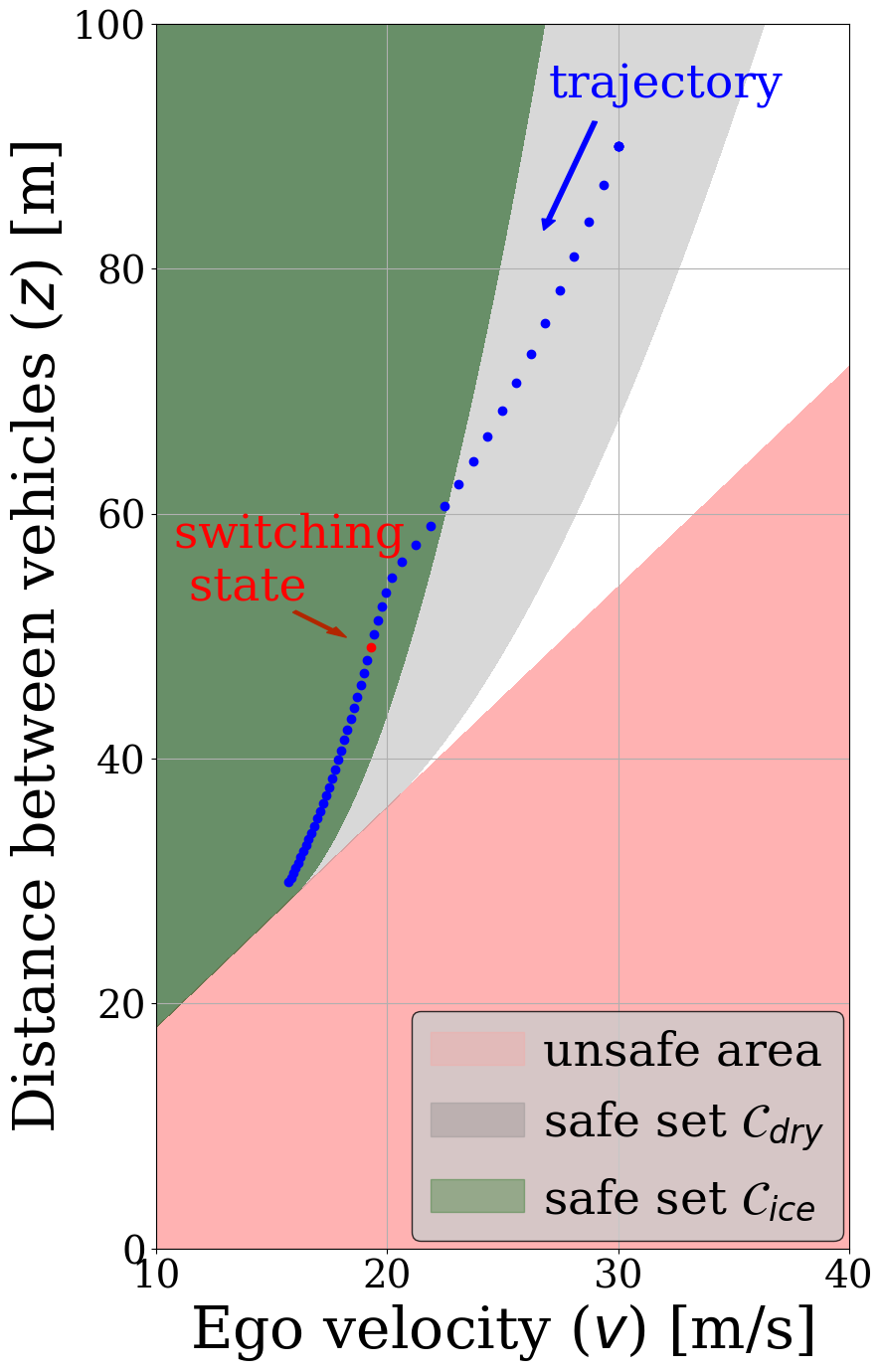}
    \caption{\scriptsize{Global CBF}}
    \label{fig:acc-global}
  \end{subfigure}
  \caption{Trajectories of our approach, switch-unaware CBF approach, and global CBF approach. The unsafe area (red) is defined by the safety constraint function $c(x)$.}
\end{figure}

In our simulation, the ego car is expected to drive with a desired speed $v_d$ (where $v_d>v_0$) while maintaining a safe distance with the leading car.
The safety specification is defined by the constraint function $c(x)=d-T_h\cdot v$.
The CBF for dry road is 
$h_{dry}(x)=d-T_hv-\frac{(v_0-v)^2}{2c_{dry}g},$
and $h_{ice}(x)$ is defined similarly by replacing $c_{dry}$ with $c_{ice}$ for ice road.

We compare our proposed approach with the switch-unaware method, in which $h_{dry}$ is applied as the safety filter while on dry road and $h_{ice}$ on the ice road, and also compare with global CBF method, where $h_{global}=h_{ice}$ is applied for both modes. 
For our approach, we first refine $h_{dry}$ to consider safe switching from dry road to ice road, and obtain $h_{dry, ice}$, which is applied as the safety filter while on dry road. The CBF is switched to $h_{ice}$ after the dynamical mode has switched. Our result is shown in Fig.~\ref{fig:ours-good-final}, in which we can notice that the switching state (red point) is in the safe set of the new dynamics ($\mathcal{C}_{ice}$ in this example).
Thus, the safety will not be violated after switching to $h_{ice}$. 
However, as it can be observed in Fig.~\ref{fig:baseline-bad-final}, in the switch-unaware controller, the switching point is not in $\mathcal{C}_{ice}$ and safety is violated after switching.
Also, global CBF method (Fig.~\ref{fig:acc-global}) is safe but more conservative than our approach since it decreases ego velocity more and will finally reach the destination later.

\subsection{Dubins car collision avoidance}
\begin{figure}
    \centering
    \includegraphics[width=0.4\textwidth]{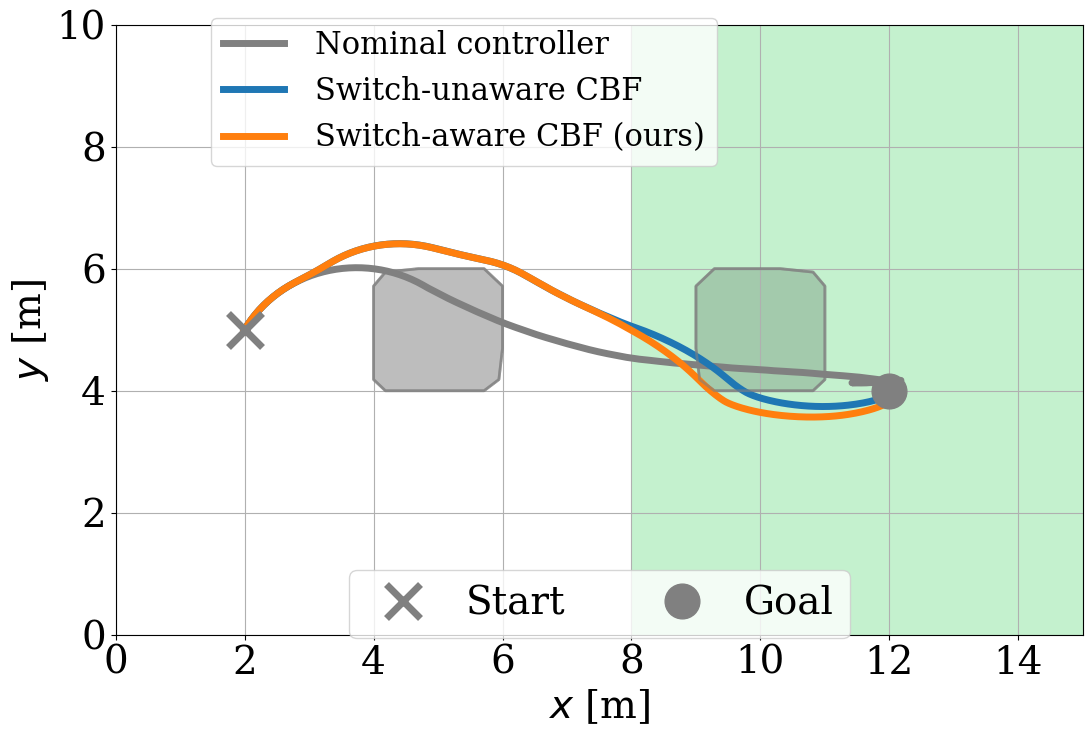}
    \caption{Dubins car is reaching a goal and avoiding two obstacles. The white and green regions are dry and wet surfaces, respectively. Gray boxes are obstacles. Nominal controller provides reference trajectory for CBF-based approaches. Global CBF method is not applicable in this case.}
    \label{fig:dubins-car}
\end{figure}

Consider an extended Dubins car model below, where $x, y$ are position, $\theta$ is heading, and $v$ is speed.
\begin{align}
    &\dot{x}=vcos(\theta), \dot{y}=vsin(\theta), \dot{v}=a, \dot{\theta}=\omega.
\end{align}
The control input includes the acceleration $a$ and angular velocity $\omega$.
The objective of the Dubins car is to navigate to reach a goal while avoiding obstacles, as shown in Fig.~\ref{fig:dubins-car}.
However, there are two different road surfaces in this task: dry road (white region on the left side of Fig.~\ref{fig:dubins-car}) and wet road (green region on the right side of Fig.~\ref{fig:dubins-car}).
We have two local CBFs $h_{dry}$ and $h_{wet}$ for them, respectively.
For different surfaces, the Dubins car has different control bounds.

As a comparison, we show trajectories of 3 different approaches in Fig.~\ref{fig:dubins-car}.
To demonstrate the performance of the CBF-based safety filters, a reference trajectory (gray) is generated without considering the obstacles, so it is unsafe.
With the switch-unaware CBF approach, i.e., applying the original local CBF for each mode as safety filters, the trajectory (blue) turns to be unsafe after switching to the ice road.
Finally, our approach refines $h_{dry}$ and obtains $h_{dry, wet}$ to ensure safe switching from dry surface to wet surface, and $h_{dry, wet}$ and $h_{wet}$ are applied as safety filters before and after switching respectively.
Our trajectory (orange) is safe.
Global CBF approach is not applicable in this example because dry road mode and wet mode have different safe regions.

%% file: conclusion.tex
\section{Conclusion and Discussion}\label{sec:conclusion}
In this work, we formulate and solve the safety control problem for hybrid systems using multiple local CBFs.
We discover that the safety of hybrid systems can be jeopardized even if all flow modes can be independently safe through control.
Thus, we propose to refine the initial local CBFs to avoid any  unsafe switching regions.
Finally, we can obtain safety guarantees under refined multiple local CBFs.

We open up a new research direction for CBFs-based safety control for hybrid systems and there are many interesting future directions. 
For example, we are working on extentions to other classes of hybrid models such as hybrid systems with non-deterministic or stochastic jumps.

\section{Acknowledgment}
The authors thank Hardik Parwana for initial discussions. 